\documentclass[aps,prb,superscriptaddress,showpacs]{revtex4}%
\usepackage{amsmath}
\usepackage{graphicx}
\usepackage{amsfonts}
\usepackage{amssymb}%
\begin{document}
\title{Macroscopic Quantum Tunneling in \textquotedblleft Small" Josephson Junctions
in magnetic field}
\author{Yu.N.Ovchinnikov}
\affiliation{Max-Plank Institute for Physics of Complex Systems, Dresden, D-01187 Germany
and Landau Institute for Theoretical Physics, RAS, Chernogolovka, Moscow
District, 142432 Russia}
\author{A. Barone}
\affiliation{Dipartimento Scienze Fisiche, Universita' di Napoli \textquotedblleft Federico
II" and Coherentia-INFM, CNR, piazzale Tecchio, 80, 80125, Napoli, Italy}
\author{A.A.Varlamov}
\affiliation{Coherentia-INFM, CNR, Viale del Politecnico 1, I-00133 Rome, Italy and
Materials Science Division, Argonne National Laboratory, 9700 S.Cass Avenue,
Argonne Il 60439, USA}
\date{\today}

\begin{abstract}
We study the phenomenon of macroscopic quantum tunneling (MQT) in small
Josephson junctions (JJ) with an externally applied magnetic field. The latter
results in the appearance of the Fraunhofer type modulation of the current
density along the barrier. The problem of MQT for a point-like JJ is reduced
to the motion of the quantum particle in the washboard potential. In the case
of a finite size JJ under consideration, this problem corresponds to a MQT in
potential which itself, besides the phase, depends on space variables.
Finally, the general expression for the crossover temperature $T_{0}$ between
thermally activated and macroscopic quantum tunneling regimes and the escaping
time $\tau_{\mathrm{esc}}$ have been calculated.

\end{abstract}
\maketitle

The paramount importance of macroscopic quantum tunneling lies in the role of
such a process in a variety of physical phenomena, which span from the
cosmological implications of the false \textquotedblright vacuum" to recent
studies on BEC systems. Great attention deserves in this context the
occurrence of MQT in Josephson junctions \cite{1} observed both in low $T_{c}$
\cite{2} and high $T_{c}$ \cite{3} structures. Usually the phenomenon of MQT
is considered in a \textquotedblleft small" JJ, i.e. in the assumption that
the characteristic size $L$ of the junction is much smaller than all other
related parameters of the problem (some exceptions can be found in
experimental \cite{a} and theoretical papers \cite{b} ). Such parameters are
the Josephson penetration depth $\lambda_{J}=\left(  \hbar c^{2}/8\pi
ej_{1}\ell_{\mathrm{eff}}\right)  ^{1/2}$\ and the characteristic length
$L_{H}=\ell_{H}^{2}/\ell_{\mathrm{eff}}$, with $\ell_{H}=\left(  \hbar
c/eH_{\mathrm{ext}}\right)  ^{1/2}$ as the standard quantum mechanical
magnetic length. The external magnetic field $H_{\mathrm{ext}}$ is supposed to
be directed along the plane of junction. The critical current density $j_{1}%
$\ of the JJ will be defined explicitly below. The effective length
$\ell_{\mathrm{eff}}$ depends on the relation between the thickness
$d_{\left(  i\right)  }$ $(i=L,R)$ of the superconductive electrodes and the
magnetic penetration depth $\lambda_{\left(  i\right)  }$ of the bulk
superconductor materials, which the electrodes are made of. In the limiting
cases one can find \cite{EST64,K65,ISS67,LO68}
\begin{equation}
\ell_{\mathrm{eff}}=\left\{
\begin{tabular}
[c]{ll}%
$\lambda_{\left(  1\right)  }+\lambda_{\left(  2\right)  }+d_{\mathrm{ox}},$ &
$\lambda_{\left(  i\right)  }\ll d_{\left(  i\right)  }$\\
$d_{\left(  1\right)  }+d_{\left(  2\right)  }+d_{\mathrm{ox}},$ & $d_{\left(
i\right)  }\ll\lambda_{\left(  i\right)  }$%
\end{tabular}
\ \ \ \ \ \ \ \ \ \right.  . \label{leff}%
\end{equation}

The finite value of the junction length $L$ becomes important even for
$L<\lambda_{J}.$ Such a situation takes place when the magnetic field is
applied to the junction what results in appearance of the Fraunhofer
diffraction pattern $I_{c}$ vs $H_{\mathrm{ext}}$. In our analysis we will
suppose that $L\ll\lambda_{J}$ but will not impose any restrictions on the
relation between $L$ and $L_{H}.$

The effect of magnetic field on the properties of JJ leads to three different
manifestations. First of all magnetic applied filed suppresses the value of
the order parameter $\Delta$ and changes the frequency dependence of the
single-particle density of states $\rho\left(  \omega\right)  $ (formally, the
presence of magnetic field changes the frequency dependence of the Green
functions) \cite{LO72} . Then, its presence leads to the spacial modulation of
the current flowing through the junction \cite{EST64,K65,ISS67,LO68,BP82}.
Finally, the magnetic field affects on the probability of the electron
tunneling through the barrier. Since the thickness of insulating barriers in
the case of JJ is very small (about $10\mathring{A}$) this effect we will
neglect in our discussion.

Let us start our analysis from the suppression of the order parameter $\Delta$
and changes of the frequency dependence of the Green functions by the magnetic
field. The Green functions integrated over the energy variable ( $\alpha
,\beta$) for both electrodes can be found from the equation \cite{LO72}:
\begin{equation}
\alpha\Delta-\beta\omega+\frac{\emph{D}}{2}\left\{  \alpha\left(
\frac{\partial}{\partial\mathbf{r}}-2ie\mathbf{A}\right)  ^{2}\beta-\beta
\frac{\partial^{2}}{\partial\mathbf{r}^{2}}\alpha\right\}  =\alpha\beta\Gamma,
\label{GF1}%
\end{equation}
with the normalization condition $\alpha^{2}\left(  \mathbf{r,}\omega\right)
+\beta^{2}\left(  \mathbf{r,}\omega\right)  =1.$ We choose the vector
potential gauge in each electrode in the form $\mathbf{A=(}Hz,0,0\mathbf{),}$
where the direction of $\ \widehat{z}$ axis is supposed to be perpendicular to
the junction surface and the magnetic field $\mathbf{H}$ is directed along the
$\widehat{y}$-axis. The coordinate $z$\ in each electrode changes in the
symmetric interval $z\in\lbrack-d/2,d/2],$ while the coordinate along the
barrier $x\in\lbrack-L/2,L/2].$\ In view of the smallness of the JJ length one
can see that the spacial dependence of $\Delta$\ and Green functions\ can be
neglected. Averaging the Eq.(\ref{GF1}) over the space variable one can
determine the frequency dependence of the averaged Green functions
($\overline{\alpha}\left(  \omega\right)  ,\overline{\beta}\left(
\omega\right)  $)%

\begin{equation}
\overline{\alpha}_{\left(  i\right)  }\Delta_{\left(  i\right)  }%
-\overline{\beta}_{\left(  i\right)  }\omega=\overline{\alpha}_{\left(
i\right)  }\overline{\beta}_{\left(  i\right)  }\left(  \Gamma_{\left(
i\right)  }+\emph{D}_{\left(  i\right)  }e^{2}H_{\mathrm{ext}}^{2}d_{\left(
i\right)  }^{2}/6\right)  , \label{GF3}%
\end{equation}
where $\Gamma_{\left(  i\right)  }=\tau_{s\left(  i\right)  }^{-1}$ is the
inverse electron spin-flip scattering time.

The critical current density $J_{1}\left(  H\right)  $ can be determined from
the Ambegaokar-Baratoff type formula \cite{BP82}:%

\begin{equation}
J_{1}\left(  H\right)  =\frac{\pi}{eR_{\mathrm{N}}S}T\sum_{\omega}%
\overline{\beta}_{\left(  1\right)  }\left(  \omega\right)  \overline{\beta
}_{\left(  2\right)  }\left(  -\omega\right)  ,\qquad\omega\equiv\omega
_{n}=2\pi T\left(  n+\frac{1}{2}\right)  ,
\end{equation}
where $R_{N}$ \ is the junction resistance in normal state and $S$ is the
junction area. Thus the effect of magnetic field results just in the
renormalization of the critical current density. The latter becomes essential
when $\emph{D}_{\left(  i\right)  }e^{2}H_{\mathrm{ext}}^{2}d_{\left(
i\right)  }^{2}\sim\Delta.$

Let us discuss of the second external magnetic field manifestation on the
properties of a JJ, related to the space modulation of current density in it.
In the case of a \textquotedblleft point-like" JJ the value $J_{1}$ determines
the height of the potential barriers and hence the escaping time for the MQT.
In the case of a small JJ of finite size the current density is space
modulated by the external magnetic field and the escaping time\ calculation
becomes more complicated. Indeed, while for the point contact it is reduced to
the motion of the quantum particle in the washboard potential in the case
under consideration the problem corresponds to the tunneling of a lengthy
object in the potential which itself, besides the phase, depends on space variables.

To calculate the escaping time $\tau_{\mathrm{esc}}$ we need to write down the
functional of effective action $A\left[  \varphi\right]  $, what can be done
in this case basing on the results of \cite{LO83a,LO83b}:
\begin{align}
A\left[  \varphi\right]   &  =\frac{1}{S}\int_{-1/2T}^{1/2T}d\tau\int
d^{2}\mathbf{r}\left\{  \frac{C}{2e^{2}}\left(  \frac{\partial\varphi\left(
\mathbf{r},\tau\right)  }{\partial\tau}\right)  ^{2}-\frac{\hbar}{e}%
I\varphi\left(  \mathbf{r},\tau\right)  -\frac{\pi\hbar}{2R_{\mathrm{N}}e^{2}%
}\int_{-1/2T}^{1/2T}d\tau_{1}\left[  2\sin^{2}\left(  \frac{\varphi\left(
\mathbf{r},\tau\right)  -\varphi\left(  \mathbf{r},\tau_{1}\right)  }%
{2}\right)  \right.  \right. \nonumber\\
&  \times\left.  \alpha_{L}\left(  \tau-\tau_{1}\right)  \alpha_{R}\left(
\tau-\tau_{1}\right)  +\cos\left(  \varphi\left(  \mathbf{r},\tau\right)
+\varphi\left(  \mathbf{r},\tau_{1}\right)  \right)  \beta_{L}\left(
\tau-\tau_{1}\right)  \beta_{R}\left(  \tau-\tau_{1}\right)  \right]
\nonumber\\
&  +\left.  \frac{\pi\hbar T^{2}}{R_{\mathrm{sh}}e^{2}}\int_{-1/2T}%
^{1/2T}d\tau_{1}\frac{\sin^{2}\left(  \frac{\varphi\left(  \mathbf{r}%
,\tau\right)  -\varphi\left(  \mathbf{r},\tau_{1}\right)  }{2}\right)  }%
{\sin^{2}\left[  \pi T\left(  \tau-\tau_{1}\right)  \right]  }+\frac{\hbar
^{2}c^{2}S}{8\pi e^{2}\ell_{\mathrm{eff}}}\left(  \frac{\partial\varphi\left(
\mathbf{r},\tau\right)  }{\partial\mathbf{r}}-\frac{e\ell_{\mathrm{eff}}%
}{\hbar c}\left[  \mathbf{H}_{\mathrm{ext}}\times\mathbf{n}\right]  \right)
^{2}\right\}  . \label{Aphi}%
\end{align}

Here $C$ is the junction capacity, $I$\ is the full current flowing through
the junction, $R_{\mathrm{sh}}$\ is the shunt resistance. The integrals are
carried out over the imaginary time and the junction area. The first term
corresponds to the kinetic energy of the junction. The second and fourth terms
describe the contribution of the potential energy to the action. Let us stress
that the third term, corresponding to the capacity renormalization appears
only beyond the quasi-classical \ approximation. The action accounting for all
these terms was already found in \cite{LO83a,LO83b} in the case of microscopic
study of the Josephson current decay in point-like junction. The last term in
Eq.( \ref{Aphi}) accounts for the finite size of the junction and the magnetic
field contribution to the effective action. The motion in the classically
permitted region is described by the same functional Eq.( \ref{Aphi}) in the
assumption $\tau=\tau_{1}$ what eliminates the third term. The further
variation procedure of \ Eq.( \ref{Aphi}) over $\varphi$ results in the
getting of the quasi-classical equations of motion \cite{EST64,K65,ISS67,LO68}.

We are looking for the escape time $\tau_{\mathrm{esc}}$ with the accuracy up
to the pre-exponential factor. The exponential factor in it is determined by
the extremal value of the functional Eq.( \ref{Aphi}) found at the class of
solutions periodical in imaginary time $\tau$ with the period $1/T.$ The
function $\varphi\left(  x,\tau\right)  $ at which the extremum of functional
Eq.( \ref{Aphi}) is realized (the saddle point of functional ) we will search
in the form%
\begin{equation}
\varphi\left(  x,\tau\right)  =\varphi_{0}-\frac{x}{L_{H}}+\widetilde{\varphi
}\left(  x,\tau\right)  =\frac{1}{2}\Phi\left(  x\right)  +\widetilde{\varphi
}\left(  x,\tau\right) , \label{phisaddle}%
\end{equation}
where $\varphi_{0}=const,$ is the phase difference at the junction in the
absence of magnetic field, while the function $\widetilde{\varphi}\left(
x,\tau\right)  $ satisfies the boundary condition \cite{LO68}
\begin{equation}
\left.  \frac{\partial\widetilde{\varphi}}{\partial x}\right\vert
_{x=-L/2}=\left.  \frac{\partial\widetilde{\varphi}}{\partial x}\right\vert
_{x=L/2}=0. \label{bound}%
\end{equation}

The experimentally realistic case corresponds to the current close to its
critical value $I_{c}$. This is why we restrict our consideration namely by
the condition $I\rightarrow I_{c}$. Let us note that in this case the third
term in Eq.( \ref{Aphi}) can be excluded from consideration by the simple
renormalization of the junction capacity $C\rightarrow C^{\ast},$ where%
\begin{equation}
C^{\ast}=C+\frac{\pi\hbar}{2R_{\mathrm{N}}}\int_{-\infty}^{\infty}%
\frac{d\omega}{2\pi}\left(  \frac{\partial\alpha_{L}}{\partial\omega}\right)
\left(  \frac{\partial\alpha_{R}}{\partial\omega}\right)  .
\end{equation}

The substitution of Eq.( \ref{phisaddle}) to the Eq.( \ref{Aphi}) with the
further expansion of the cosine over $\widetilde{\varphi}\left(
x,\tau\right)  $ in the fourth term leads to the cancellation of the linear in
$\widetilde{\varphi}$ second term. Since the function $\widetilde{\varphi}$
has to be periodical in imaginary time $\tau$ with the period $1/T$ we present
it in the form of the Fourier series. This makes the effective action (
\ref{Aphi}) to become the functional of the Fourier coefficients $b_{n}$ (let
us note that $b_{n}\left(  x\right)  =b_{-n}\left(  x\right)  $):%
\begin{align}
A\left[  b_{n}\right]   &  =\frac{\hbar J_{1}}{eT}\int d^{2}\mathbf{r}%
\sum_{n=-\infty}^{\infty}\left\{  \left[  \cos\left[  \Phi\left(  x\right)
\right]  +\zeta\left(  n^{2}+Q_{\mathrm{sh}}|n|\right)  \right]  b_{n}%
^{2}\right. \nonumber\\
&  \left.  -\frac{2}{3}\sin\left[  \Phi\left(  x\right)  \right]  \left(
\sum_{k=-\infty}^{\infty}b_{k}b_{k+n}\right)  b_{n}+\lambda_{J}^{2}\left(
\frac{\partial b_{n}}{\partial x}\right)  ^{2}\right\}  , \label{Abn1}%
\end{align}
where we have introduced the notations $\zeta=2\pi^{2}T^{2}C^{\ast}/\left(
e\hbar J_{1}S\right)  $\ and the Q-factor $Q_{\mathrm{sh}}^{-1}=2\pi
TR_{\mathrm{sh}}C^{\ast}/\hbar.$The further variation of the functional
(\ref{Abn1}) over $b_{n}$ results in the system of equations:%
\begin{equation}
-\lambda_{J}^{2}\frac{\partial^{2}b_{n}}{\partial x^{2}}+\left[  \cos\left[
\Phi\left(  x\right)  \right]  +\zeta\left(  n^{2}+Q_{\mathrm{sh}}|n|\right)
\right]  b_{n}-\sin\left[  \Phi\left(  x\right)  \right]  \left[  (
2-\delta_{n,0})b_{0}b_{n}+2\sum_{k=1}^{\infty}b_{k}b_{n+k}+\sum_{k=1}%
^{n-1}b_{k}b_{n-k}\right]  =0 \label{sys2}%
\end{equation}
Let us recall that we will look for the solution of this system with the
boundary conditions ( \ref{bound}). The analogous system of equations for the
point junction was solved for the cases of low and high shunt resistance
\cite{LO83a} and also close to the temperature $T_{0}$ corresponding to the
crossover between the thermally activating (Arrenius law) and quantum regimes
in tunneling through the barrier.

In order to calculate the escape time we need to know the value of the
effective action ( \ref{Aphi}) at the saddle point, i.e. at the optimal
trajectories $b_{n}\left(  x\right)  ,$ being solutions of the system (
\ref{sys2}). The related expression can be considerably simplified,
subtracting from the functional Eq. ( \ref{Abn1}) the Eq. ( \ref{sys2})
multiplied by corresponding $b_{n}\left(  x\right)  :$%

\begin{equation}
\widetilde{A}\left[  b_{n}\right]  =\frac{\hbar J_{1}}{3eT}\int d^{2}%
\mathbf{r}\sin\left[  \Phi\left(  x\right)  \right]  \left[  b_{0}^{3}%
+6b_{0}\sum_{n=1}^{\infty}b_{n}^{2}+4\sum_{n=1}^{\infty}b_{n}\sum
_{k=1}^{\infty}b_{k}b_{n+k}+2\sum_{n=1}^{\infty}b_{n}\sum_{k=1}^{n-1}%
b_{k}b_{n-k}\right]  . \label{areduced}%
\end{equation}

We consider the junction of a small enough size: $L\ll\lambda_{J}.$ Thus\ the
system Eq. (\ref{sys2}) acquires the large parameter $\lambda_{J}^{2}/L^{2}%
\gg1$. Consequently the functions $b_{n}\left(  x\right)  $\ weakly vary with
the change of its argument and can be represented in the form:%

\begin{equation}
b_{n}=\left\langle b_{n}\right\rangle +\widetilde{b}_{n},\qquad|\widetilde
{b}_{n}|\ll|\left\langle b_{n}\right\rangle |, \label{16}%
\end{equation}
where $\left\langle ...\right\rangle $ indicates the average over the
coordinate $x$ along the junction. The corresponding equation for
$\widetilde{b}_{n}$\ takes the form%
\begin{equation}
\lambda_{J}^{2}\frac{\partial^{2}\widetilde{b}_{n}}{\partial x^{2}%
}=\left\langle b_{n}\right\rangle \sin\left(  2\varphi_{0}\right)  \sin\left(
\frac{2x}{L_{H}}\right)  . \label{17}%
\end{equation}
Its solution satisfying the boundary conditions ( \ref{bound}) is%

\begin{equation}
\widetilde{b}_{n}=\frac{L_{H}^{2}}{4\lambda_{J}^{2}}\left\langle
b_{n}\right\rangle \sin\left(  2\varphi_{0}\right)  \left[  \left(  \frac
{2x}{L_{H}}\right)  \cos\left(  \frac{L}{2L_{H}}\right)  -\sin\left(
\frac{2x}{L_{H}}\right)  \right]  . \label{18}%
\end{equation}
What concerns the equations for the average values $\left\langle
b_{n}\right\rangle $ they can be obtained directly from the Eq.( \ref{sys2})
by the method of subsequent approximations.

Now we can calculate the escape time $\tau_{\mathrm{esc}}$ near the crossover
temperature $T_{0},$ taking into account the effects of the current density
coordinate dependence, due to the applied magnetic field. All coefficients
$\left\langle b_{n}\right\rangle $ with $n\geq1$ are small and quickly
decrease as $n$ increases. It is why at temperatures\ close to $T_{0}$ in Eq.
( \ref{areduced}) one can restrict consideration by the contribution of
$b_{0},$ $b_{1}$ and $b_{2}\left(  \sim b_{1}^{2}\right)  .$

The crossover temperature $T_{0}$ is characterized by the fact that below it
the quantum tunneling through barrier becomes more probable than the thermally
activated escape process. Formally this temperature is the bifurcation point
of the system ( \ref{sys2}) when the first non-zero solutions for $n\geq1$
appear. Above it \ exists the only, trivial, solution is $b_{0}=const,$ what
results in the action ( \ref{Aphi}) $\widetilde{A}\left[  \widetilde{\varphi
}\right]  =\Delta U/T$ corresponding to the Arrenius law for the escape time.
Below $T_{0}$ the non-zero solutions $b_{n}$ appear, what leads to the
appearance of the imaginary time dependence of $\widetilde{\varphi}$ , the
action ( \ref{Aphi}) changes its form and,consequently, the temperature
dependence of the escape time $\tau_{\mathrm{esc}}\sim\exp\left[  -A\left(
\varphi_{0},L\right)  \right]  $ \ changes qualitatively. After some
cumbersome but straightforward calculations one can find the quadratic
equation determining the crossover temperature $T_{0}:$%

\begin{equation}
\frac{2\pi^{2}C^{\ast}T_{0}^{2}}{\hbar e}+\frac{\pi T_{0}}{R_{\mathrm{sh}}%
e}=I_{c}\sin\left(  2\varphi_{0}\right)  \left(  \frac{L_{H}}{L}\right)
\sin\left(  \frac{L}{L_{H}}\right)  \Upsilon\left(  \varphi_{0},L\right)  ,
\label{quadratic}%
\end{equation}
where $I_{c}=J_{1}(H)S$ is the value of junction's critical current, while
\begin{equation}
\Upsilon\left(  \varphi_{0},L\right)  =\cot\left(  2\varphi_{0}\right)
+\sin\left(  2\varphi_{0}\right)  \left(  \frac{L_{H}}{2\lambda_{J}}\right)
^{2}\left[  \frac{3}{2}\cos\left(  \frac{L}{L_{H}}\right)  -\left(  \frac
{L}{L_{H}}\right)  \frac{1+\frac{1}{2}\cos\left(  \frac{2L}{L_{H}}\right)
}{\sin\left(  \frac{L}{L_{H}}\right)  }\right]  .
\end{equation}
In the case of a point junction the right-hand-side of Eq. ( \ref{quadratic})
reduces to $\sqrt{I_{c}^{2}-I^{2}}$.

In vicinity of $T_{0}$ the essential contribution to Eq. ( \ref{areduced})
\ give only \ the terms with $b_{0},b_{1}$ and $b_{2}.$ Their average values
can be found from Eq.( \ref{16})-( \ref{18}). We present here only%

\begin{equation}
\left\langle b_{1}\right\rangle ^{2}=\frac{\Upsilon\left(  \varphi
_{0},L\right)  \zeta\sin\left(  2\varphi_{0}\right)  \left(  \frac{L_{H}}%
{L}\right)  \sin\left(  \frac{L}{L_{H}}\right)  -\zeta^{2}\left(
Q_{\mathrm{sh}}+1\right)  }{2\sin^{2}\left(  2\varphi_{0}\right)  \left(
\frac{L_{H}}{L}\right)  ^{2}\sin^{2}\left(  \frac{L}{L_{H}}\right)  \left[
\frac{2}{Q_{\mathrm{sh}}+1}-\frac{1}{Q_{\mathrm{sh}}+3}\right]  }.
\end{equation}
what finally permits to find
\begin{equation}
\widetilde{A}\left[  \varphi\right]  =\frac{\hbar J_{1}S}{3eT}\sin\left(
2\varphi_{0}\right)  \left(  \frac{L_{H}}{L}\right)  \sin\left(  \frac
{L}{L_{H}}\right)  \left[  \Upsilon^{3}\left(  \varphi_{0},L\right)
-\frac{6\left\langle b_{1}\right\rangle ^{4}}{\zeta}\sin\left(  2\varphi
_{0}\right)  \left(  \frac{L_{H}}{L}\right)  \sin\left(  \frac{L}{L_{H}%
}\right)  \left(  \frac{2}{Q_{\mathrm{sh}}+1}-\frac{1}{Q_{\mathrm{sh}}%
+3}\right)  \right]  .\nonumber
\end{equation}

One can see that the effect of magnetic field in a finite size junction, even
in the case $L\ll\lambda_{J},$\ does not reduce simply to the renormalization
of the value of $I_{c}.$ Indeed, both characteristic parameters of the
problem, $T_{0}$ and the escape time $\tau_{\mathrm{esc}},$ turn out to be the
functions of two parameters: $I/I_{c}$\ and $L/L_{H}.$ Let us stress that when
the current $I$ is very close to its critical value $I_{c}$ the approximation
(\ref{16}) for the functions $b_{n}\left(  x\right)  $ becomes incorrect, and
this region has to be studied separately. These results are of great interest
for the whole scenario of quantum coherence, as for instance in the context of
MQT in Josephson structures based on d-wave superconductors \cite{3,a}

Yu.N.Ovchinnikov acknowledges the financial support of the grant
RFBR-06-01-90507-BNTS-a and Russian Foundation of Basic Research.

\end{document}